\documentclass[aps,prb,twocolumn,superscriptaddress,showkeys]{revtex4}
\usepackage{graphicx}
\usepackage{amsmath}
\usepackage{longtable}

\begin{document}

\title{%
Phase-sensitive Harmonic Measurements of Microwave Nonlinearities in Cuprate Thin Films
}

\author{Dragos I. Mircea}
\affiliation { 
San Jose Research Center, Hitachi Global Storage Technologies, 3403 Yerba Buena Rd., San Jose, CA 95135
}
\author{Hua Xu}
\affiliation{ 
Center for Nanoscale Science and Technology, National Institute of Standards and Technology, Gaithersburg, MD 20899
}
\author{Steven M. Anlage}
\affiliation{ 
Center for Nanophysics and Advanced Materials, Department of Physics, University of Maryland, College Park, MD 20742
}

\date{\today}

\begin{abstract}
Investigations of the intrinsic electromagnetic nonlinearity of superconductors give insight into the fundamental physics of these materials. Phase-sensitive third-order harmonic voltage data $\tilde{u}_{3f}=|u_{3f}|exp(\!i\phi_{3f}\!)$ are acquired with a near-field microwave microscope on homogeneous YBa$_2$Cu$_3$O$_{7-\delta}$ (YBCO) thin films in a temperature range close to the critical temperature T$_c$. As temperature is increased from below T$_c$, the harmonic magnitude exhibits a maximum, while the phase, $\pi/2$ in the superconducting state, goes through a minimum. It is found that samples with doping ranges from near optimal ($\delta=0.16$) to underdoped ($\delta=0.47$) exhibit different behavior in terms of both the harmonic magnitude and phase. In optimally-doped samples, the harmonic magnitude reaches its maximum at a temperature $T_M$ slightly lower than that associated with the minimum of phase $T_m$ and drops into the noisefloor as soon as $T_m$ is exceeded. In underdoped samples $T_M$ is shifted toward lower temperatures with respect to $T_m$ and the harmonic voltage magnitude decreases slower with temperature than in the case of optimally-doped samples. A field-based analytical model of $\tilde{u}_{3f}$ is presented, where the nonlinear behavior is introduced as corrections to the low-field, linear-response complex conductivity. The model reproduces the low-temperature regime where the $\sigma_2$ nonlinearity dominates, in agreement with published theoretical and experimental results. Additionally the model identifies $T_m$ as the temperature where the order parameter relaxation time becomes comparable to the microwave probing period and reproduces semi-quantitatively the experimental data. 
\end{abstract}
\maketitle

\section{
INTRODUCTION
}

Recently, a number of experiments have shown evidence of unusual properties above the superconducting transition temperature in under-doped cuprate superconductors. Observations include a significant Nernst effect in the pseudo-gap phase of La-Sr-Cu-O, suggesting the existence of vortex excitations \cite{Ong}. Diamagnetic response above T$_c$ has also been observed in under-doped Bi-Sr-Ca-Cu-O \cite{Ong2}. These properties have been generally interpreted in terms of a superconducting state with non-zero superfluid density but dominated by strong phase fluctuations of the order parameter. Such a state should have interesting nonlinear response characterized by persistence of superconducting nonlinearities above T$_c$, as reported before with scalar nonlinearity measurements \cite{Lee-PRB1}. The present study extends these results by measuring the complex harmonic voltage developed by under-doped cuprate superconductors when temperature is varied through the transition temperature.

Traditionally, the microwave nonlinear response of superconductors has been investigated by using resonator techniques where the superconducting sample is subject to high microwave magnetic fields, thus making the nonlinear effects measurable \cite{Amato,Trunin}. The experimental data have been interpreted by using various time-dependent versions of the Ginzbug-Landau theory to estimate the order parameter relaxation time in the superconducting state \cite{Gorkov}. After the discovery of high-T$_c$ superconductors, the prospect of using these materials in microwave filters for the wireless industry has renewed interest in the microwave nonlinear response. In the more recent treatments, the nonlinear effects are introduced as corrections to the complex conductivity and are evaluated by using a microscopic approach in the zero-frequency limit \cite{Dahm-Scalapino,Yip-Sauls}. The DC treatment is legitimate for the range of temperatures typical for the operation of high-T$_c$ superconducting filters (below T$_c$) where the superconducting order parameter reacts almost instantaneously (compared to the period of the microwave excitation) to the applied field and the field screening is provided by the superfluid. The resulting field (or current density)-dependent conductivity is used as an input parameter for calculations of circuit elements in lumped-element approximations of the superconducting transmission lines and resonators \cite{Dahm-Scalapino}. Theoretical studies addressing the operation of high-T$_c$ superconducting resonators have shown that for temperatures significantly below T$_c$, the dominant nonlinear mechanism in these devices has an inductive origin due to the suppression of superfluid density by the current (or applied magnetic field) \cite{Dahm-Scalapino}. 

Resonator techniques have provided experimental support for the Nonlinear Meissner Effect (NLME) at low temperatures in d-wave superconductors \cite{Oates2008} (enhanced by the presence of nodes of the order parameter on the Fermi surface as shown in the theoretical works of Xu, Yip and Sauls \cite{Yip-Sauls} and Dahm and Scalapino \cite{Dahm-Scalapino}) as well as close to T$_c$ where the superfluid density is very sensitive to external perturbations \cite{Amato,Trunin}. Despite their success, the resonator techniques measure parts of the sample that are often less than ideal, such as patterned edges, or natural edges and corners of single crystals. Such experiments do not provide information about the local properties of the samples and also usually do not provide phase information of the nonlinear response. This issue is relevant especially for high-T$_c$ materials whose properties may vary on very short length scales due to their short coherence lengths. 

To overcome this limitation, a nonresonant local near-field microwave technique has been created to make spatially-resolved studies of nonlinear response of superconductors \cite{Pestov,Lee-APL,Lee-PRB1}. This experimental approach is highly sensitive to nonlinear effects close to T$_c$ and provides a high spatial resolution dictated by the geometrical dimensions of the sensing element, as demonstrated in harmonic measurements above an artificially-created grain boundary \cite{Lee-APL,Lee-PRB2}. The basic idea is to excite a highly localized current distribution at frequency $f$ on the surface of a homogeneous unpatterned superconductor. Due to the nonlinear electrodynamic processes, harmonic ($2f$, $3f$, etc.) signals are created in the material and collected by the sensing element. The harmonic data measured at T$_c$ on YBa$_2$Cu$_3$O$_{7-\delta}$ (YBCO) thin films has been interpreted in the framework of a Ginzburg-Landau-type model where the nonlinear source is the magnetic field(current)-dependent superfluid density n$_S(T,J)$ \cite{Lee-PRB1}. The current-suppressed superfluid density leads to an enhancement of the penetration depth $\lambda(T,J)$, and consequently of the kinetic inductance, which in turn, leads to odd higher-order harmonics of inductive origin. The proposed model describes accurately the measured data (temperature-dependent third-order harmonic scalar power P$_{3f}$(T)) in optimally-doped samples. This approach also largely avoids issues of nonlocality that are exacerbated by current build-up at patterned edges \cite{Dahm-Scalapino,Oates2008}.

However, in underdoped samples the current-dependent superfluid density n$_S(T,J)$ acting alone as an inductive nonlinear source cannot explain the observed harmonic data. More specifically, the measured harmonic response P$_{3f}(T)$ does not turn off at temperatures above the independently determined T$_c$ as expected from the model, but exhibits a tail extending significantly above T$_c$ in the pseudogap regime \cite{Lee-PRB1}. It is this high-temperature behavior of P$_{3f}(T)$, including its origin and doping dependence, that prompted the present study. In addition, many theoretical predictions of interesting electrodynamic properties of the pseudogap exist in the literature \cite{Anderson,Tan,Mishonov-PRB,Dorsey,VarlamovReggiani,Puica,Kurin}, and should be investigated. By employing a novel experimental technique, the harmonic response of cuprate thin films has been investigated at temperatures close to T$_c$, where not only the magnitude of the complex harmonic voltage/power is measured, as in previous work, but also its harmonic phase. This new experimental capability also motivates a new field-based, finite-frequency model to explain novel features observed mainly in the harmonic phase experimental data. The model includes the microwave skin depth screening in the electrodynamics of the superconducting state close to T$_c$, where the superfluid density is suppressed and consequently its field screening is compromised.

The paper is organized as follows: section \ref{experimental} describes the new experimental set-up and the samples used in this study. Emphasis is placed on presenting in detail the acquisition and data processing methodology. Section \ref{model} presents a new theoretical model aimed at evaluating the complex harmonic voltage measured with the experimental set-up presented in the previous section. By using general electromagnetic theory and making minimal assumptions about the nature of nonlinear effects, the model provides predictions in the temperature regime where $\sigma_1\ll\sigma_2$ and $\sigma_1\approx\sigma_2$. These limiting cases are discussed in section \ref{analysis} in conjunction with our experimental data. The model reproduces in a semi-quantitative fashion some features observed in the experimental data as temperature is increased toward T$_c$. The harmonic data acquired on samples with various doping levels are discussed also in section \ref{analysis}. The Discussion and Conclusions sections describe the main features of the data and the predictions of the model. Some deficiencies of the model are pointed out and suggestions are made for future work. 

\section{
EXPERIMENTAL SET-UP AND SAMPLES
} \label{experimental}

\begin{figure}
\includegraphics[width=0.95\columnwidth]{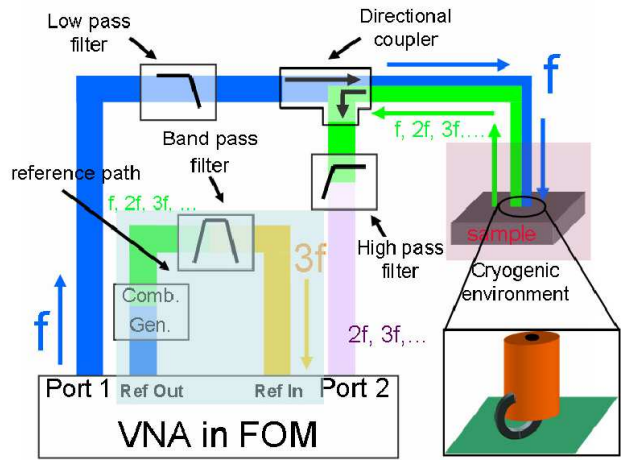}
\caption{\label{VNA_FOM_setup}(Color online) The experimental set-up for the phase-sensitive harmonic measurements. A signal at frequency $f$ is generated by the internal VNA source, low-pass filtered and delivered to the sample inside a cryogenic environment. The microwave signal reflected by the sample and containing harmonics of the incident power is high-pass filtered to remove the fundamental $f$ and measured on VNA port 2. For phase-sensitive detection a reference circuit, converts some microwave power at frequency $f$ into power at $3f$ by means of a harmonic generator and two band-pass filters, and feeds the resulting $3f$ signal back into the VNA as reference signal. Inset: drawing of the magnetic loop probe above the sample.}
\end{figure}

The objective of the experiment (shown in Fig.~\ref{VNA_FOM_setup}) is to locally stimulate a homogeneous superconducting thin film with microwave currents and measure the resulting nonlinear response. The microwave excitation is provided by the internal source of a vector network analyzer, VNA, (Agilent model number E8364B) on port 1 at a fixed frequency $f\approx 6.5$ GHz in the continuous-wave mode, low pass filtered (to eliminate harmonics) and coupled to the sample by means of a magnetic loop probe. The probe is built by using commercially-available coaxial cable (UT034) where the inner conductor has been soldered to the outer one \cite{Pestov,Lee-APL,Lee-PRB1}. This results in a semicircular loop with inner radius of roughly $165 \mu m$ and an outer one of $365 \mu m$ (see inset of Fig.~\ref{VNA_FOM_setup}). The loop has been further mechanically polished at the outer radius  in order to bring the microwave current, flowing in a thickness dictated by the skin depth at the inner radius of the loop, closer to the sample ($\sim 100 \mu$m), thus improving the loop-to-sample coupling and allowing the operation of the apparatus at lower input power levels. The signal originating from the sample comes back through the probe, is high-pass filtered to suppress the microwave power at the fundamental frequency $f$ and examined with the VNA in the frequency-offset mode (VNA-FOM) by tuning the receiver on port 2 in a narrow frequency range (1 Hz) centered on the harmonic of interest ($3f$ in this case).

The sample is placed in a cryogenic environment whose temperature is controlled between 78 and 100 K with an accuracy of 0.1 K and for each temperature a trace is acquired from the VNA-FOM and stored on a computer for further analysis. The samples are unpatterned homogeneous c-axis oriented YBa$_2$Cu$_3$O$_{7-\delta}$ (YBCO) thin films deposited on SrTiO$_3$ or NdGaO$_3$ substrates by pulsed laser deposition (PLD). The oxygen content of the samples has been adjusted by annealing in various oxygen pressures and at different temperatures in the PLD chamber resulting in critical temperatures T$^{AC}_c$ in the range of 52 to 90 K, as evaluated from AC susceptibility measurements (performed at 120 kHz). Despite their small thickness ($\sim 50$ nm) the sample superconducting quality is very good as revealed by the narrow peaks of the temperature-dependent imaginary part of the magnetic susceptibility whose full width at half maximum $\delta T_c^{AC}$ are given in Table~\ref{table1}.

\begin{table}[h]
\begin{center}
\begin{tabular}{|c|c|c|c|c|c|}
\hline
Sample & T$_c^{AC}$ [K] & $\delta T_c^{AC}$ [K] & $7-\delta$ & $\Delta T_{M,m}$ [K] & Substrate\\
\hline
S1 &88.9&0.3&6.84&0.2&NdGaO$_3$\\
\hline
S2 &86.6&1.0&6.82&0.5&NdGaO$_3$\\
\hline
S3 &74.0&0.9&6.76&0.4&NdGaO$_3$\\
\hline
S4 &62.0&0.55&6.69&0.6&SrTiO$_3$\\
\hline
S5 &52.0&1.1&6.53&1.0&SrTiO$_3$\\
\hline
\end{tabular}
\caption{\label{table1}Sample properties: critical temperature $T^{AC}_c$ and transition width $\delta T_c^{AC}$ determined from AC susceptibility measurements, the doping level $7-\delta$ estimated from $T^{AC}_c$, the difference between the temperatures where the extreme values of the harmonic phase and magnitude occur, $\Delta T_{M,m}$, and the sample substrate.}
\end{center}
\end{table}
 
After a power calibration, performed according to the manufacturer's instructions \cite{AgilentFOM}, the VNA-FOM measures the absolute power level at frequency $3f$, $P_{3f}$, incident on port 2 similar to a spectrum analyzer ("spectrum analyzer mode"). 

\begin{figure}
\includegraphics[width=1\columnwidth]{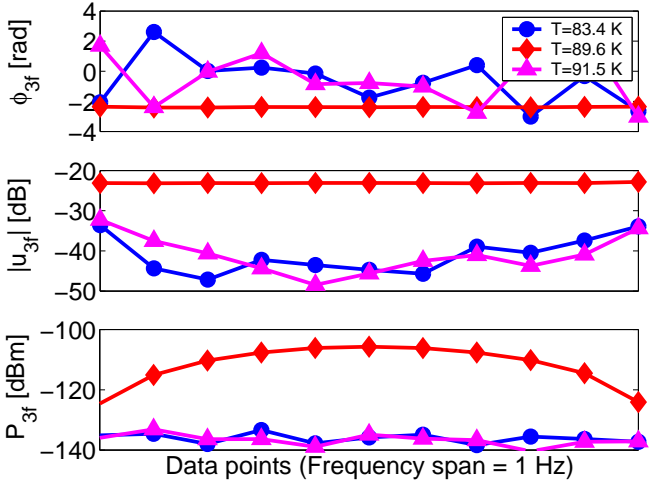}
\caption{\label{VNA_FOM_traces}(Color online) Examples of VNA-FOM traces acquired on a YBCO thin film (S1) in a frequency range centered on $3f=19.47$ GHz. Top and middle plot: the phase and magnitude of $\tilde u_{3f}(T)$ acquired in a phase-sensitive measurement; bottom plot: absolute harmonic power data $P_{3f}(T)$, acquired in "spectrum analyzer mode".}
\end{figure}

To perform phase-sensitive detection of the $3f$ harmonic voltage incident on port 2, the VNA-FOM requires a \textit{reference signal} at the same frequency as the signal to be analyzed. For the measurements reported here, the reference signal is provided by an additional microwave circuit, called the \textit{reference path} (see Fig.~\ref{VNA_FOM_setup}) that converts some microwave power at frequency $f$ generated by the internal VNA source into microwave power at frequency $3f$. The fundamental $f$ is fed into a comb (harmonic) generator (Herotek, model number GCA 2026A-12) followed by two band-pass filters designed to suppress (attenuation of 80 dB) the fundamental and all harmonics except for $3f$, thus resulting in a clean $3f$ signal that serves as reference, $U_{3f}^{ref}$, for the phase-sensitive harmonic detection. The VNA-FOM traces represent the complex ratio of the voltage from the sample (whose temperature $T$ is varied inside the cryostat), to that from the reference path, $\tilde u_{3f}(T)=U_{3f}^{sample}(T)/U_{3f}^{ref}$ at the plane of the VNA's port 2, evaluated at the frequency points scanned by the VNA-FOM receiver within the 1 Hz span window centered on $3f$ ("vector signal analyzer mode"). Since during an experiment the reference path and most of the microwave circuit are at room temperature (only about 10 cm of coaxial cable is inside the cryostat, however not in physical contact with the cold plate), it is legitimate to attribute the temperature dependence of the measured relative harmonic voltage $\tilde u_{3f}(T)$ entirely to the temperature-dependent nonlinear effects in the sample. Since the microwave circuit is operated at low microwave power ($\sim 0$ to $+9$ dBm at VNA port 1), the background noise at $3f$ is dominated by the intrinsic noise of the VNA ($\sim -140$ dBm when $P_{3f}$ is measured without an input on VNA port 2). This is shown in Fig.~\ref{VNA_FOM_traces}, bottom plot, where the nonlinear signal from the sample are at the noisefloor (T=83.4 and 91.5 K).

The VNA-FOM is not designed for absolute phase harmonic measurements, but for relative ones since the VNA-FOM only indicates how the harmonic phase \textit{changes} as the sample properties change from one temperature to another, provided that the reference signal $U_{3f}^{ref}$ is stable during the measurements. Consequently, the temperature-dependent VNA-FOM phase data $\phi_{3f}(T)=\Phi_{3f}^{sample}(T)-\Phi_{3f}^{ref}-\Phi_{offset}$, are offset by a temperature-independent unknown amount ($\Phi_{3f}^{ref}+\Phi_{offset}$) originating from the phase winding in the components of the microwave circuit (coaxial cables and filters) and the phase relationship between the fundamental and the harmonics generated by the comb generator. The phase shift $\Phi_{3f}^{ref}+\Phi_{offset}$ is evaluated by using the predictions of the theoretical model presented in Section \ref{model} for the limiting case of low temperatures ($T\ll T_c$) where the harmonic phase is $\pi/2$, with a minimum of assumptions and in agreement with other experimental observations \cite{Booth2005}.

To increase the signal-to-noise ratio of the VNA-FOM traces, 8-10 averages were performed on the VNA before transferring the trace to the computer. Such averaged traces acquired at three representative temperatures (below, above and around $T_c^{AC}$) are shown in Fig.~\ref{VNA_FOM_traces}: the complex phase and magnitude of the harmonic voltage (upper and middle plots) and the absolute harmonic power levels (bottom plot). The 1 Hz span $P_{3f}$ VNA-FOM traces look similar to small-span traces acquired with a spectrum analyzer: at T=89.6 K (close to $T_c^{AC}$) the temperature-dependent harmonic power $P_{3f}(T)$ reaches its maximum in agreement with other authors \cite{Amato,Trunin,Pestov,Lee-PRB1}, while at T=83.4 K and T=91.5 K (below and above $T_c^{AC}$ respectively) the traces are flat and at the noisefloor ($-135$ to $-140$ dBm). Complex $\tilde u_{3f}(T)$ VNA-FOM averaged traces acquired at T=83.4 K and T=91.5 K, exhibit a large scatter both in phase and magnitude, while $|\tilde u_{3f}(T)|$ reaches its maximum at T=89.6 K, in agreement with the $P_{3f}$ traces. 

\begin{figure}
\includegraphics[width=1\columnwidth]{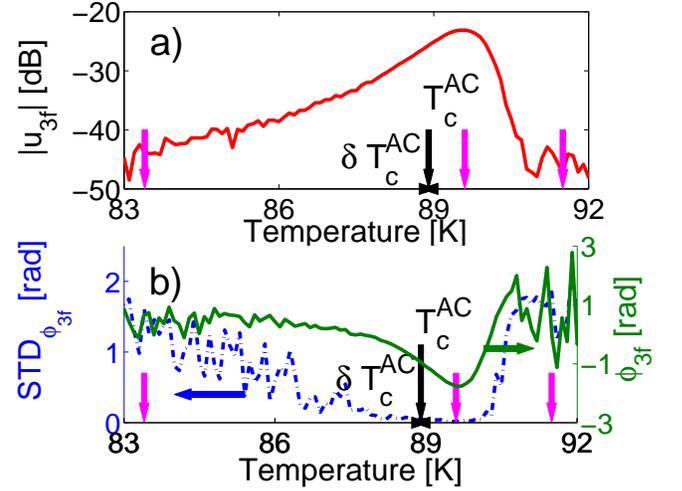}
\caption{\label{VNA_FOM_data}(Color online) Phase-sensitive harmonic data acquired on a YBCO thin film (S1). a): temperature-dependent magnitude, $|u_{3f}(T)|$. b): temperature-dependent phase ($\phi_{3f}(T)$ green solid line) and the standard deviation of the 11-point traces acquired at each temperature (blue dashed line). Magenta arrows show: T=83.4, 89.6 and 91.5 K.}
\end{figure}

To extract the temperature dependence of the relative harmonic voltage $\tilde u_{3f}(T)$ from the VNA-FOM traces, 1 to 3 central points from each trace (see Fig.~\ref{VNA_FOM_traces}) are averaged, and the resulting complex magnitude and phase are plotted vs. temperature in Fig.~\ref{VNA_FOM_data}. To quantify the data spread in a $\tilde u_{3f}(T)$ VNA-FOM trace and the reliability of averaging the complex traces, the standard deviation of the phase data $STD_{\phi_{3f}}$ is evaluated from each of the averaged 1 Hz span traces and represented together with the phase data in Fig.~\ref{VNA_FOM_data}. The temperature-dependent $STD_{\phi_{3f}}$ can be used to select a temperature range where $\phi_{3f}(T)$ data can be considered reliable by imposing that $STD_{\phi_{3f}}$ does not exceed a certain threshold. The temperatures associated with the traces from Fig.~\ref{VNA_FOM_traces} are indicated in Fig.~\ref{VNA_FOM_data} with magenta arrows together with $T_c^{AC}$ and $\delta T_c^{AC}$ from AC susceptibility measurements.  

The samples have been measured by using various input frequencies (6.45 to 6.55 GHz) and power levels ($0$ to $+9$ dBm) and with the microwave probe placed at several locations above the samples, all with consistent results. For the range of microwave input power levels employed in this work, the harmonic data suggest that the microwave probe does not induce a significant amount of heating in the sample surface, which would be indicated by a shift of the maximum of $|\tilde u_{3f}(T)|$ and of the minimum of $\phi_{3f}(T)$ respectively to lower temperatures when the microwave power is increased. 

Qualitatively, the magnitude of the harmonic voltage, $|u_{3f}(T)|$, reaches a maximum at a temperature $T_M$ close to $T_c^{AC}$ in agreement with results from the literature where $P_{3f}(T)$ is reported to reach a maximum \cite{Amato,Trunin,Pestov,Lee-PRB1}. One novelty of the results presented here is that the complex phase decreases smoothly and reaches a minimum at $T_m$ as the temperature is increased. For the samples investigated here a consistent trend has been observed: for near-optimally doped samples the temperatures associated with the two extrema of magnitude and phase almost coincide, while for underdoped samples the harmonic phase tends to reach its minimum at higher temperatures ($T_m>T_M$). To quantify this trend, $\Delta T_{M,m}=T_m-T_M$ has been evaluated for all samples from measurements at various input frequencies and microwave power levels and is given in Table \ref{table1}. There is a general trend of increasing $\Delta T_{M,m}$ with increased underdoping. 

\section{
MODEL 
} \label{model}

The recent theoretical treatments of microwave nonlinear effects in superconductors are mostly restricted to resonant configurations and their equivalent lumped-element circuit approximations \cite{Dahm-Scalapino}. Due to the non-resonant nature of the near-field microwave microscope employed in this work and since the sample is not part of a transmission line structure, a field-based analytical model is more appropriate to capture the essential physics. Such an approach has been proposed for temperatures below T$_c$, where the authors considered only the nonlinear effects caused by the field dependence of the imaginary part of the complex conductivity \cite{Pestov}. 

The analytical model presented in this section is an extension of that proposed by Pestov and co-workers \cite{Pestov} at finite frequencies in an attempt to explain the temperature dependence of the harmonic phase data acquired with the near-field microwave microscope described in the previous section. The goal of the model is to provide an expression for the complex-valued temperature-dependent harmonic voltage $U_{3f}^{sample}(T)$ induced in the near-field antenna by the microwave screening current distribution from the nonlinear superconducting sample. This is achieved in three steps: first, the magnetic vector potential created by the excitation current in the microwave probe, $A_f$, is calculated at the sample surface, then by using a nonlinear generalization of the constitutive London equation for superconductors, the harmonic content of the screening current induced in the sample is evaluated, and in the third step the harmonic voltage induced in the near-field probe is found.

In order to preserve a higher level of generality, the nonlinear effects are introduced here as phenomenological corrections to both the real and imaginary parts of the low-power, linear-response complex conductivity of the sample $\tilde{\sigma}=\sigma_1-i\sigma_2$, where $\sigma_{1,2}$ are positive definite ($\sigma_{1,2}\ge 0$):
\begin{equation}\label{sigmaNL}
\sigma_{1,2}(T,A_f)=\sigma_{1,2}\left (\!1\pm\frac{A_f^2}{A^2_{1,2}}+ \cdots \!\!\right ), A_f \ll A_{1,2}
\end{equation}
$A_f$ is the vector potential associated with the microwave excitation at the fundamental frequency $f$ and the temperature-dependent nonlinear vector potential scales $A_{1,2}$ quantify the nonlinear effects in the real and imaginary components of the complex conductivity, respectively. We assume local electrodynamics in this model. Note that $A_{1,2}$ can model a wide variety of nonlinear sources, including the nonlinear Meissner effect, vortex motion, weak links, etc. These corrections are valid when $A_f \ll A_{1,2}$, similar to other phenomenological descriptions of nonlinear effects in the literature \cite{Pestov,Booth2005,Lee-PRB1}. This treatment also implicitly assumes that the complex conductivity of the superconductor reacts instantaneously to changes of the probing vector potential $A_f$. At temperatures very close to $T_c$, where the condition $A_f \ll A_{1,2}$ might be violated and higher order terms should be included in the expansion (\ref{sigmaNL}), or when order parameter relaxation times become comparable to the microwave period, the present formalism might not be applicable. In a qualitative picture the superfluid density $n_S$ is suppressed by the microwave excitation ("-" sign in Eq.(\ref{sigmaNL}) for $\sigma_2$) and "converted" into normal fluid ("+" sign in Eq.(\ref{sigmaNL}) for $\sigma_1$). 

\begin{figure}
\includegraphics[width=1\columnwidth]{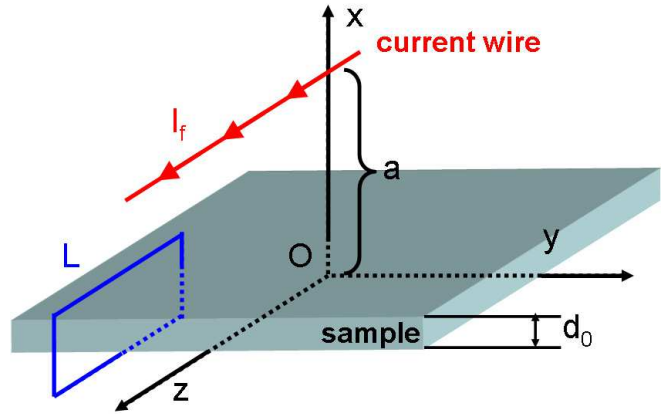}
\caption{\label{model_setup}(Color online) The geometry of the analytical problem. The near-field microwave antenna is approximated by a filamentary current wire extending infinitely in the z direction at $x=a$, $y=0$. The sample with thickness d$_0$ extends infinitely in the zOy plane and has its bottom surface at $x=0$.}
\end{figure}

For the analytical treatment of the problem, the near-field probe is approximated by a filamentary current wire parallel to the z axis located at $y=0$ and $x=a$ whose current density is modeled by the $\delta$ Dirac distribution: $j_f^{ext}(x,y)=I_f\delta(x-a)\delta(y)$, while the sample extends infinitely in the zOy plane of the Cartesian frame with the bottom surface at $x=0$ (see Fig.~\ref{model_setup}). Since the wire-to-sample separation, $a$, is much smaller than the wavelength of the microwave excitation, the magnetic vector potential $A_f$ outside the superconducting sample satisfies the equations of magnetostatics: above the sample $\nabla^2A_f=-\mu_0j_f^{ext}(x,y)$ while below the sample $\nabla^2A_f=0$. Inside the superconducting sample the spatial variation of the magnetic vector potential $A_f$ and current density in the film $j_{film}$ in the $x$ direction is governed by the length scale associated with the inverse of the wave vector $|\gamma^{-1}|$ ($\gamma^2=\lambda^{-2}+2i\delta_{sk}^{-2}(f)$ which at low temperature $T\ll T_c$ is governed by the penetration depth $\lambda$ while in the normal state by the skin depth evaluated at the probing frequency $f$, $\delta_{sk}(f)$). For details on the superconducting screening at finite frequencies, see for example the work of Clem and Coffey \cite{CoffeyClem}. Since the sample thickness is smaller than both the zero-temperature penetration depth as well as the microwave skin depth within the investigated temperature range ($d_0 \ll \lambda_0, \delta_{sk}(f)$), $A_f$ and $j_{film}$ are assumed uniform within the thickness $d_0$. At microwave frequencies the displacement current is negligible with respect to the conduction one, therefore Ampere's law in integral form for the closed loop L (shown in Fig.~\ref{model_setup}) reads:
\begin{equation}\label{Faraday_law}
B_f(x=d_0+0)-B_f(x=0-0)=\mu_0j_{film}d_0
\end{equation}
where $j_{film}d_0$ is the sheet current. In the limit $d_0\to 0$ of sample thickness much smaller than all length scales involved in the problem, and taking into account that the current density inside the film obeys London's law for finite-frequency: $j_{film}=-\gamma^2A_f/\mu_0$ (here we assume local electrodynamics), the equation for the magnetic vector potential, can be written in a closed-form for the entire space \cite{Pestov}:
\begin{equation}\label{AeqLIN}
-\nabla ^2 A_f(x,y)+\lambda_{eff}^{-1}A_f(x,y) \delta(x)=\mu_0I_f\delta(x-a)\delta(y)
\end{equation}
where $\lambda_{eff}=1/(d_0\gamma^2)$ represents a generalized finite frequency effective penetration depth. 

To integrate Eq.(\ref{AeqLIN}), the nonlinear effects in $\sigma_{1,2}$ are neglected in this step ($A_{1,2}\to \infty$) and the equation is Fourier-transformed. After solving for $A_f(k_x,k_y)$ and integrating with respect to $k_x$, $A_f(k_y)$ reads \cite{Pestov}:
\begin{equation} \label{AFourier}
A_f(k_y)= \mu_0 I_f \frac{\lambda_{eff}\exp(-|k_y|a)}{1+2\lambda_{eff}|k_y|} 
\end{equation}

For the experimental setup described here, the sample-to-wire separation, $a$ is determined by the diameter of the inner conductor of the coaxial cable ($a \sim 100 \mu$m). In the long-wavelength approximation, $a$ exceeds both length scales contained in $\lambda_{eff}$ ($\lambda$ and $\delta_{sk}(f)$) and consequently the term $2\lambda_{eff}|k_y|$ can be neglected in the denominator of Eq.(\ref{AFourier}), allowing a closed-form expression for the vector potential generated by the current wire:
\begin{equation} \label{linearA}
A_f(y)\cong -\frac{\mu_0 I_fa}{\pi d_0(a^2+y^2)}\cdot \frac{1}{\lambda^{-2}+2i\delta_{sk}^{-2}(f)} 
\end{equation}

The finite-frequency nonlinear generalization of the local London constitutive relationship $j_{film}=j_S+j_n= (\sigma_1-i\sigma_2)E=-\omega(i\sigma_1+\sigma_2)A_f$ (where the electric field $E=-\partial A_f/\partial t$ and $A_f\sim \exp(+i\omega t)$ with $\omega=2\pi f$) is obtained by replacing the linear-response complex conductivity $\sigma_{1,2}$ with its phenomenological nonlinear expressions from  Eq.(\ref{sigmaNL}):
\begin{equation} 
j_{film}\cong-\omega\sigma_2\!\!\left(\!\!1-\frac{A_f^2}{A^2_2}\!\!\right)\!\!A_f-i\omega\sigma_1\!\!\left (\!\!1+\frac{A_f^2}{A^2_1}\!\!\right)\!\!A_f
\end{equation}
or can be expressed in terms of the linear-response length scales $\lambda$ and $\delta_{sk}$:
\begin{equation} 
j_{film}\cong-\frac{1}{\mu_0\lambda^2}\!\!\left(\!\!1-\frac{A_f^2}{A^2_2}\!\!\right)\!\!A_f-\frac{2i}{\mu_0\delta_{sk}^2(f)}\!\!\left(\!\!1+\frac{A_f^2}{A^2_1}\!\!\right)\!\!A_f
\end{equation}

This approximation is valid under the limited condition $A_f\ll A_{1,2}$ and shows that the current density contains a component at frequency $f$ and another component at frequency $3f$ which represents the source of the measured harmonic voltage at frequency $3f$. The nonlinear component at frequency $3f$ in the total current density $j_{film}$, $j_{3f}$, is separated from the $A_f^3$ terms by considering the time dependence $A_f\sim \cos(\omega t)$ and using the trigonometric relation $\cos^3\omega t=(\cos3\omega t+3\cos\omega t)/4$:
\begin{equation} \label{J_NL}
j_{3f}=\frac{A_f^3}{4\mu_0}\left(\frac{1}{\lambda^2A^2_2}-\frac{2i}{\delta_{sk}^2A^2_1}\right)=\frac{\omega}{4}\left(\frac{\sigma_2}{A_2^2}-i\frac{\sigma_1}{A_1^2}\right)A_f^3
\end{equation}

The current distribution flowing in the sample and having a harmonic $3f$ time variation generates a vector potential $A_{3f}$ in the entire space and induces a voltage in the near-field probe. In order to evaluate the induced voltage at $3f$, $U_{3f}^{sample}$, one has to calculate the vector potential at the location of the wire. This is accomplished by using the equivalence principle from electromagnetism \cite{Collin} where a current with frequency $3f$ flowing through the wire $j_{3f}^{ext}=I_{3f}\delta(x-a)\delta(y)$ generates the magnetic vector potential on the sample surface given by Eq.(\ref{linearA}) with the appropriate substitution $f\to 3f$. Equivalently, a current distribution $j_{3f}$ in the sample given by Eq.(\ref{J_NL}) generates a vector potential $A_{3f}(x,y,z)$ in the entire space. The equivalence principle \cite{Collin} reads:
\begin{equation} 
\!\!\int\!\!\!\mathrm{d}V\!j_{3f}^{ext}(\!x,y,z\!)A^{3f}(\!x,y,z\!)\!=\!\!\!\int\!\!\!\mathrm{d}V\!j_{3f}(\!x,y,z\!)A_{3f}(\!x,y,z\!)
\end{equation}
with the integrals evaluated over the entire space. By using the filtering properties of the Dirac delta function and since all the $z=constant$ planes contain the same field and current configuration due to the symmetry of the problem, the vector potential at the location of the wire reads:
\begin{align} 
A^{3f}(a,0)=\frac{5}{64}\left(\!\frac{\mu_0I_f}{\pi d_0a}\!\right)^3\left(\!\!\frac{1}{\lambda^2A_2^2}-\frac{2i}{\delta_{sk}^2(f)A_1^2}\!\right) \cdot 
\nonumber \\* 
\cdot\left(\!\frac{1}{\lambda^{-2}+2i\delta_{sk}^{-2}(f)}\!\right)^3\frac{1}{\lambda^{-2}+2i\delta_{sk}^{-2}(3f)}
\end{align}
where $\delta_{sk}(f)$ and $\delta_{sk}(3f)$ represent the skin depth evaluated at frequency $f$ and $3f$ respectively. These two quantities differ by a factor of $\sqrt{3}$ in the ordinary skin effect regime, and in order to simplify the calculations, the following approximation will be used: $\delta_{sk}(3f)\approx\delta_{sk}(f)=\delta_{sk}$. 

The electric field induced in the wire at frequency $3f$, $E_{3f}=-\partial A^{3f}(a,0)/\partial t$, is used to evaluate the voltage induced in a probe of length $l_0$:
\begin{align}\label{U3fspli22}
U_{3f}^{sample}(a,0)=\frac{15\omega l_0}{64}\left(\!\frac{\mu_0I_f}{\pi d_0a}\!\right)^3\frac{\lambda^6}{A_1^2}\cdot 
\nonumber \\*
\cdot\left[\frac{2\lambda^2}{\delta_{sk}^2}+i\frac{A_1^2}{A_2^2}\right]\left[1+i\frac{2\lambda^2}{\delta_{sk}^2}\right]^{-4}
\end{align}
In terms of conductivities, the induced voltage reads:
\begin{align}\label{U3fspli22Sigma}
U_{3f}^{sample}(a,0)=\frac{15\omega l_0}{64}\left(\!\frac{I_f}{\pi d_0a\omega}\!\right)^3\frac{1}{\sigma_2^3A_1^2}\cdot 
\nonumber \\*
\cdot\left[\frac{\sigma_1}{\sigma_2}+i\frac{A_1^2}{A_2^2}\right]\left[1+i\frac{\sigma_1}{\sigma_2}\right]^{-4}
\end{align}

Equations  (\ref{U3fspli22}) and (\ref{U3fspli22Sigma}) have been deduced in an analytical, \textit{field-based} approach, as opposed to most recent models in the literature which use lumped-element descriptions for the superconducting devices operating at microwave frequencies. Several features can be noted: the harmonic voltage magnitude scales with the excitation current as $|U_{3f}^{sample}|\sim I_f^3$, the nonlinear effects are easier to measure in thin films ($|U_{3f}^{sample}|\sim d_0^{-3}$) and for small antenna-to-sample geometric separation ($|U_{3f}^{sample}|\sim a^{-3}$), all in agreement with experimental data in the literature, as well as models. 

The model provides an estimate for the complex-valued harmonic voltage induced in the near-field probe by the screening current flowing on the sample surface. Since the VNA-FOM measures the harmonic voltage from the sample with respect to that from the reference path at the plane of VNA's port 2, the measured data must be phase-shifted by an amount $\Phi_{3f}^{ref}+\Phi_{offset}$ (see Section \ref{experimental}), which is equivalent to moving the measurement plane from VNA's port 2 to the near-field antenna. Such a translation is effectively accomplished by examining the limiting case $T\ll T_c$ of Eq.(\ref{U3fspli22Sigma}) to evaluate the required amount of phase shift.

\section{
ANALYSIS
} \label{analysis}

\subsection{
Harmonic Phase vs. Temperature
} \label{phaseVsT}

Before comparing the results of the mathematical model with the experimental data, it is useful to examine the case of nonlinearity at low temperatures, which has also been investigated by other authors in a resonant configuration \cite{Booth2005}. For $T\ll T_c$, when the contribution of the normal fluid to the electrodynamics of the superconducting state is small, the induced voltage given by Eq.(\ref{U3fspli22Sigma}) can be expanded in a power series around $\sigma_1/\sigma_2=0$:
\begin{align}\label{U3fINDUCTIVE1}
U_{3f}^{sample}\left(\!\!\frac{T}{T_c}\ll 1,a,0\!\!\right)\approx\frac{15\omega l_0}{64}\left(\!\!\frac{\mu_0I_f}{\pi d_0a\omega}\!\!\right)^3\frac{1}{\sigma_2^3A_1^2}\cdot 
\nonumber \\*
\cdot\left[i\frac{A_1^2}{A_2^2}+\left(\!\!1\!+\!4\frac{A_1^2}{A_2^2}\!\right)\frac{\sigma_1}{\sigma_2}\!-\!2i\!\left(\!\!2\!+\!5\frac{A_1^2}{A_2^2}\!\right)\!\left(\!\frac{\sigma_1}{\sigma_2}\!\right)^2\!\!\!+\!\cdots\!\right]
\end{align}

Two possible scenarios emerge from this picture depending on the ratios of conductivities and that of vector potential scales. For the case $1\gg\frac{\sigma_1}{\sigma_2}\gg\frac{A_1^2}{A_2^2}$ the harmonic voltage $U_{3f}^{sample}(a,0)$ has a zero phase and depends on the $\sigma_1$ nonlinearity characterized by the vector potential scale $A_1$:
\begin{align}\label{U3fRESISTIVE}
U_{3f}^{sample}\left(\!\!\frac{T}{T_c}\ll 1,\frac{\sigma_1}{\sigma_2}\gg\frac{A_1^2}{A_2^2},a,0\!\!\right)\approx 
\nonumber \\*
\approx\frac{15\omega l_0}{64}\left(\!\!\frac{\mu_0I_f}{\pi d_0a\omega}\!\!\right)^3\!\!\frac{1}{\sigma_2^3A_1^2}\frac{\sigma_1}{\sigma_2}
\end{align}

On the other hand, for the case $\frac{\sigma_1}{\sigma_2}\ll\frac{A_1^2}{A_2^2}$ the first term in the power expansion Eq.(\ref{U3fINDUCTIVE1}) is dominant and represents a pure inductive-like nonlinear response. The harmonic voltage $U_{3f}^{sample}(a,0)$ depends only on the $\sigma_2$ nonlinearity, characterized by the nonlinear vector potential scale $A_2$, as in most of the  treatments of superconductor nonlinear response \cite{Dahm-Scalapino,Pestov,Booth2005,Lee-PRB1}:
\begin{align}\label{U3fINDUCTIVE2}
U_{3f}^{sample}\left(\!\!\frac{T}{T_c}\ll 1,\frac{\sigma_1}{\sigma_2}\ll\frac{A_1^2}{A_2^2},a,0\!\!\right)\approx 
\nonumber \\*
\approx\frac{15\omega l_0}{64}\left(\!\!\frac{\mu_0I_f}{\pi d_0a\omega}\!\!\right)^3\!\!\frac{i}{\sigma_2^3A_2^2}
\end{align}

This result does not rely on any assumptions about the temperature dependence of conductivity correction factors $A_{1,2}$ and is in good agreement with the experimental data of Booth \textit{et al.} who reported a purely inductive harmonic response in YBCO thin films at 78 K \cite{Booth2005}, well below T$_c\approx 90$ K. Additionally, theoretical models describing the operation of high-T$_c$ microwave filters at temperatures below T$_c$, predict a purely inductive nonlinear response (consistent with our case $\frac{\sigma_1}{\sigma_2}\ll\frac{A_1^2}{A_2^2}$, $\frac{A_1^2}{A_2^2}\gg 1$), however do not exclude the possibility of resistive nonlinear effects close to T$_c$ \cite{Dahm-ScalapinoComment}. Equation (\ref{U3fINDUCTIVE2}) shows that in the superconducting state where $\frac{\sigma_1}{\sigma_2}\ll 1$, the experimental setup is not sensitive to a possible $\sigma_1$ nonlinearity; moreover, even if the $\sigma_1$ nonlinearity would dominate the $\sigma_2$ one ($\frac{\sigma_1}{\sigma_2}\ll\frac{A_1^2}{A_2^2}\ll 1$), this effect would not be detectable due to the modest contribution of $\sigma_1$ in the field screening process.

The harmonic phase data reported here exhibit an almost flat plateau at low temperatures where the magnitude starts to go above the noisefloor (see, for example, Fig.\ref{VNA_FOM_data}). This observation, together with the prediction of Eq.(\ref{U3fINDUCTIVE2}) suggests that the harmonic response in that temperature range is characterized by a $\pi/2$ phase. By using this result, the relative phase data acquired by the VNA-FOM on all samples have been corrected over the entire temperature range by adding a temperature-independent phase offset (labeled $\Phi_{3f}^{ref}+\Phi_{offset}$ in section \ref{experimental}) to enforce the condition $\phi_{3f}(T\ll T_c)\approx\pi/2$. 

As temperature is increased toward T$_c$, the ratio $\sigma_1/\sigma_2$ increases and the in-phase component starts to become significant (the second term in square brackets in Eq.(\ref{U3fINDUCTIVE1}) which is real and positive) while the out-of-phase component (the difference of the first and third terms in Eq.(\ref{U3fINDUCTIVE1})) is gradually reduced. This behavior is consistent with the data from all samples, (see, for example, Fig.\ref{VNA_FOM_data}) showing that the third harmonic phase angle rotates clockwise from $\pi/2$ as $T_c$ is approached from below.

Equation (\ref{U3fspli22Sigma}) for the harmonic voltage includes the ratios $\sigma_1/\sigma_2$, $A_1^2/A_2^2$, and $A_1$, $\sigma_2$ whose temperature dependence must be known in order to model the experimental values of the harmonic voltage magnitude and phase. For a semi-quantitative discussion it is important to examine the temperature-dependent phase of the third harmonic voltage by considering only the functional dependence of the temperature-dependent terms in Eq.(\ref{U3fspli22Sigma}): 
\begin{equation}\label{U3fspli3Sigma}
U_{3f}^{sample}(a,0) \sim \frac{1}{\sigma_2^3A_1^2}\left [\frac{\sigma_1}{\sigma_2}+i\frac{A_1^2}{A_2^2}\right ]\left [1+i\frac{\sigma_1}{\sigma_2}\right ]^{-4}
\end{equation}
The behavior of the harmonic voltage in the complex plane is dominated by the last term in Eq.(\ref{U3fspli3Sigma}) due to its 4$^{th}$ power. For superconductors, in a mean-field approximation, the ratio of conductivities $\sigma_1/\sigma_2$ is essentially zero at $T\ll T_c$ and generally increases towards infinity as the temperature approaches $T_c$; therefore the temperature dependence of the complex phase associated with the last term in Eq.~(\ref{U3fspli3Sigma}) is 0 at low temperature and executes a full 360 degrees clockwise rotation in the complex plane as temperature is increased toward $T_c$. This prediction for the sense of rotation is in agreement with the experimental data up to the temperature T$_m$ where $\phi_{3f}$ goes through its minimum. 

\begin{figure}
\includegraphics[width=1\columnwidth]{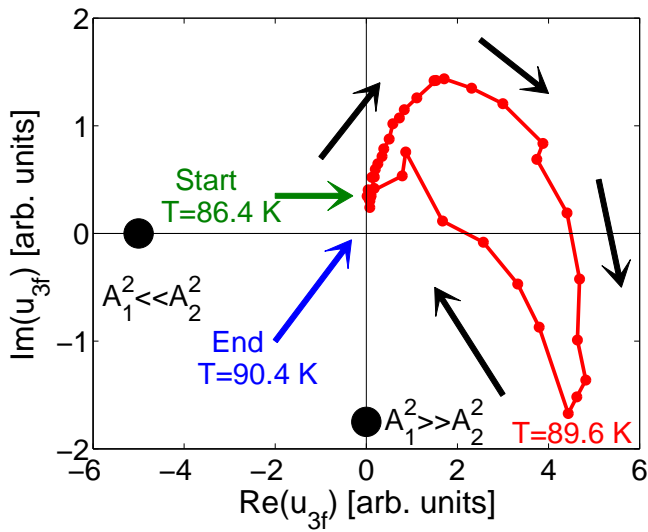}
\caption{\label{S1_complex_plane}(Color online) Temperature-dependent phase-sensitive third-order harmonic voltage data $\tilde u_{3f}$ acquired on a YBCO (S1) thin film represented in the complex plane. The arrows indicate the evolution of the complex data as temperature increases from Start(T=86.4 K) to End(T=90.4 K). Only the low-noise data are presented, here in arbitrary units. The dots represent the predicted phase at a temperature for which $\sigma_1=\sigma_2$ and either $A_1^2\ll A_2^2$ or $A_1^2\gg A_2^2$.}
\end{figure} 

The harmonic voltage $U_{3f}^{sample}(a,0)$ given by Eq.(\ref{U3fspli22Sigma}) can be expanded in a power series around $\sigma_1/\sigma_2=1$ and the first two terms are:
\begin{align}\label{U3fspli3SigmaONE}
U_{3f}^{sample}\left(\!\!\frac{\sigma_1}{\sigma_2}\approx 1,a,0\!\!\right)\!\approx\!\frac{15\omega l_0}{256}\left(\!\!\frac{I_f}{\pi d_0a\omega}\!\!\right)^3\!\!\frac{1}{\sigma_2^3A_1^2}\cdot 
\nonumber \\
\left\lbrace\!-1\!-\!i\frac{A_1^2}{A_2^2}\!+\!\!\left[1\!\!-\!\!2\frac{A_1^2}{A_2^2}\!+\!2i\!\left(\!\!1+\frac{A_1^2}{A_2^2}\!\right)\right]\!\!\left(\!\!\frac{\sigma_1}{\sigma_2}-1\!\!\right)\!\!+\!\cdots\right\rbrace
\end{align}
In the limiting case $\sigma_1=\sigma_2$ the complex harmonic voltage $U_{3f}^{sample}(a,0)$ lies in the 3$^{rd}$ quadrant of the complex plane with negative real and imaginary parts. To check this theoretical prediction, the experimental data shown in Fig.\ref{VNA_FOM_data} have been represented in the complex plane as Re($u_{3f}$) vs. Im($u_{3f}$) in Fig.\ref{S1_complex_plane}, after offsetting the phase data to enforce the condition $\phi_{3f}\approx \pi/2$ at the lowest temperature where the signal-to-noise ratio is acceptable. Also in Fig.\ref{S1_complex_plane} the two extreme cases of $A_1^2\ll A_2^2, \sigma_1=\sigma_2$ and $A_1^2\gg A_2^2, \sigma_1=\sigma_2$ have been represented as two dots. Common to all samples from Table \ref{table1}, the harmonic phase data exhibits a non-monotonic behavior, decreasing from $\pi/2$ at low temperatures, reaching a minimum inside the 4$^{th}$ quadrant at a temperature T$_m$, and increasing back toward $\pi/2$. The experimental data acquired with all the samples from Table \ref{table1} do not reach the 3$^{rd}$ quadrant of the complex plane, as Eq.(\ref{U3fspli3SigmaONE}) predicts. Note that the model does not take into account the finite order parameter relaxation time which, close to the critical temperature (i.e. in the regime $\sigma_1\approx\sigma_2$), becomes comparable to the microwave probing period. 

Experimental data acquired at low frequency (1 kHz) by Mawatari and co-workers \cite{Mawatari}, although in a different experimental configuration that enhances the electromagnetic response associated with vortex motion, exhibits a similar trend. The 3$^{rd}$ order harmonic phase goes through a minimum of roughly $-\pi/2$ when superconductivity is gradually weakened by the application of an external magnetic field, whereas for the data presented here the superconductivity is suppressed by increasing temperature. The microwave harmonic measurements presented here have been performed on a time scale roughly 7 orders of magnitude shorter than that employed in the work of Mawatari \textit{et al.} \cite{Mawatari}, thus they are more prone to explore the regime where the superconducting order parameter cannot oscillate in phase with the probing field. This could prevent the harmonic phase data from reaching the 3$^{rd}$ quadrant of the complex plane. The minimum of the harmonic phase observed in the microwave harmonic data could indicate the onset of the regime where the dynamics of the superconducting order parameter becomes slower than the period of the probing electromagnetic field. To account for this regime and to explain third-order harmonic data acquired in gapless superconductors, Amato and McLean \cite{Amato} proposed a correction term  $(1-2i\omega\tau_R)^{-1}$ to the microwave harmonic field $B_{3f}$, where $\tau_R(T/T_c)$ is the order parameter relaxation time that diverges at T$_c$. Such a correction term gives rise to a counter-clockwise rotation of the harmonic voltage in the complex plane, in agreement with our experimental data at temperatures above T$_m$.

\subsection{
Harmonic Signal vs. Doping
} \label{dopingDependence}

\begin{figure}
\includegraphics[width=1\columnwidth]{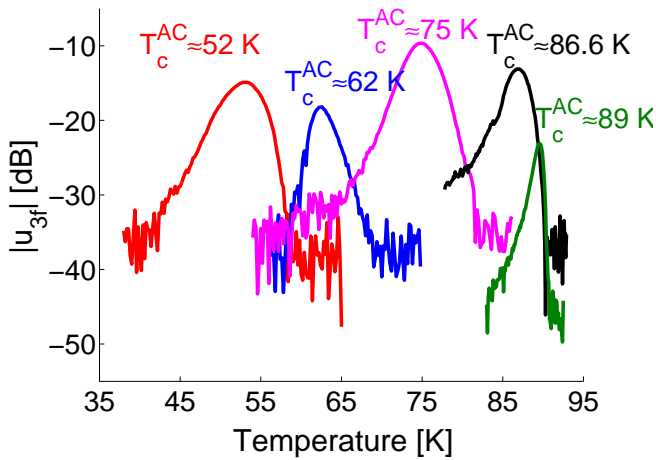}
\caption{\label{comparison_MAG}(Color online) Comparison of the third-order harmonic voltage magnitude data versus temperature acquired on the samples listed in Table \ref{table1}.}
\end{figure}

\begin{figure}
\includegraphics[width=1\columnwidth]{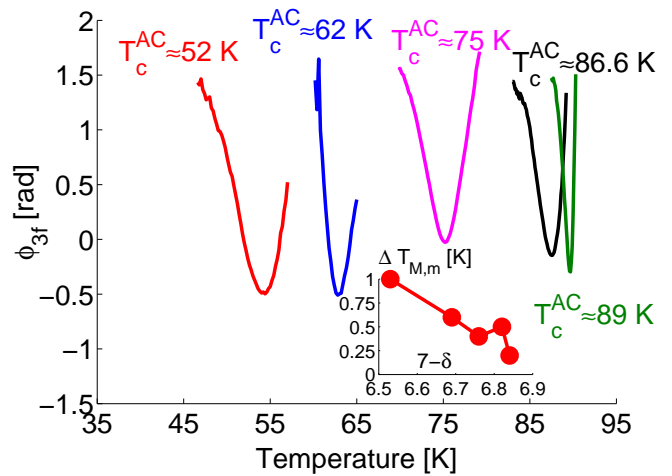}
\caption{\label{comparison_PHASE}(Color online) Comparison of the phase of the third-order harmonic voltage data versus temperature acquired on the samples listed in Table \ref{table1}. Inset: doping dependence of $\Delta T_{M,m}$.}
\end{figure}

A comparison of the magnitude and phase data acquired on all samples from Table \ref{table1} is shown in Fig.\ref{comparison_MAG} and \ref{comparison_PHASE}. Only the relatively low-noise data have been presented in these figures: for magnitude data, the noisefloor is given by the sensitivity of the VNA-FOM, while for phase data only the temperature ranges where the phase standard deviation $STD_{\Phi_{3f}}$ is less than 0.2 radians have been selected. 

The examination of both Fig.~\ref{comparison_MAG} and Fig.~\ref{comparison_PHASE} reveals that the maximum of the harmonic voltage magnitude occurs at a temperature $T_M$, lower than that associated with the minimum of the complex phase, T$_m$. The difference between these two temperatures, denoted $\Delta T_{M,m}=T_m-T_M$, follows a consistent trend as indicated in Table~\ref{table1} and shown in the inset of Fig.~\ref{comparison_PHASE} (i.e. increasing in the more underdoped samples), and is not correlated with the broadening of the superconducting transition in underdoped samples due to the annealing process (see Table \ref{table1} for $\delta T_c^{AC}$). This feature has been observed in all measurements despite the different input power levels, probing frequency or microwave antenna location above the sample. Unfortunately, due to the unavailability of $\sigma_{1,2}$ and $A_{1,2}$ theoretical temperature- and doping dependences, the observed trend of $\Delta T_{M,m}$ with doping could not be accounted for by the model. However, future theoretical models could be checked against this experimental observation.

The harmonic voltage magnitude data from Fig.\ref{comparison_MAG} show that in near optimally-doped samples, the harmonic voltage drops sharply into the noisefloor at a temperature slightly above $T_m$, while in underdoped samples, the $|u_{3f}|$ peak occurs at a temperature below $T_m$ and the $|u_{3f}|$ temperature dependence extends more above $T_m$. This general trend is in agreement with the observation of Lee \textit{et al.,} who examined the microwave harmonic power $P_{3f}$ reflected by YBCO thin films with various doping levels \cite{Lee-PRB1}. An interesting feature can be noticed in the third-order harmonic voltage phase data (Fig.~\ref{comparison_PHASE}): in optimally-doped samples when the magnitude $|u_{3f}(T)|$ reaches the noisefloor the harmonic phase recovers its value from the superconducting state ($+\pi/2$) after exhibiting a minimum at $T_m$, whereas in the most underdoped samples, the recovery is not complete. 

Some features in Fig.\ref{comparison_MAG} and Fig.\ref{comparison_PHASE} can be explained qualitatively by using the argument of Amato and McLean \cite{Amato} regarding the regime where the superconducting order parameter cannot react instantaneously to the microwave excitation. The weakening of the harmonic phase reversal and the broader extent of the harmonic magnitude observed in our harmonic data measured in the underdoped samples points to a slower divergence of the order parameter relaxation time in underdoped samples compared to their optimally-doped counterparts. Such a scenario is compatible with stronger and longer-lived superconducting fluctuations in underdoped cuprates. The extension of the nonlinear response above T$_c$ is consistent with Anderson's picture in which the pseudogap phase has the electrodynamic properties of a superconductor, but with a current-current correlation function that decays with a finite time $\tau$, and a diamagnetic susceptibility that is nonlinear \cite{Anderson}.

\section{
DISCUSSION
}\label{discussion}

Overall, the model offers a unified picture of microwave nonlinear effects originating from both the real and the imaginary parts of the conductivity. The model reproduces semi-quantitatively the trends observed in both the magnitude and phase of the harmonic voltage acquired with the phase-sensitive nonlinear near-field setup. The lack of theoretical models for the temperature- and doping dependence of $\sigma_{1,2}$ and the divergence of the order parameter relaxation time close to the critical temperature hinders a more detailed comparison of data with the predictions of the model and the extraction of the $A_{1,2}$ temperature- and doping dependence.
 
The analytical treatment points out the significance of the $\sigma_1$ nonlinearity in addition to that associated with $\sigma_2$. Most of the recent treatments of the nonlinear effects in cuprates close to the critical temperature T$_c$ ignore the influence of the normal fluid and consequently assign the observed nonlinear effects entirely to the $\sigma_2$ nonlinearity \cite{Pestov, Lee-PRB1}. Well below T$_c$, the observed nonlinear behavior is due to the superfluid, as shown in the theoretical works of Dahm and Scalapino \cite{Dahm-Scalapino}, confirmed by the experimental work of Booth \cite{Booth2005} and also shown in our model for the case of low temperatures ($\sigma_1\ll\sigma_2$). However, the microscopic model of Dahm and Scalapino does not exclude the possibility that in close proximity to T$_c$ nonlinear mechanisms due to the normal fluid might become important \cite{Dahm-ScalapinoComment}.

In the model, the relaxation time of the order parameter was assumed much shorter than the microwave period, i.e. the order parameter reacts instantaneously with changes in the external probing field. This assumption is valid only up to temperatures very close to $T_c$ \cite{Gorkov}. The new data presented here correlated with those of Mawatari \textit{et al}\cite{Mawatari} acquired at much lower excitation frequency suggest that neglecting the dynamics of the superconducting order parameter restricts our analysis to temperatures below T$_m$ where the phase of the harmonic voltage reaches its minimum in the 4$^{th}$ quadrant of the complex plane.

The major benefit of the experimental technique comes from the localized nature of the microwave excitation and the ability to measure the temperature-dependent complex harmonic voltage. Therefore, the nonlinear microwave response can be investigated in as-grown superconducting samples, free from potential defects caused by patterning. Additionally, the harmonic response can be measured at various locations in a homogeneous sample, thus ensuring that the response does not depend on location or some peculiar feature of the sample, such as edges, corners, grain boundaries or defects of fabrication\cite{AnlageBookChapter}. 

The ability to measure both the magnitude and the phase of the harmonic voltage allows a more complete description of nonlinear effects as a function of doping level. The experimental data show that the maximum of harmonic voltage magnitude and the minimum of phase occur at slightly different temperatures, $T_M$ and $T_m$ respectively. The doping dependence of the nonlinear response has been quantified by defining $\Delta T_{M,m}=T_m-T_M$ and monitoring its variation with doping level $7-\delta$. In almost optimally-doped samples the harmonic voltage magnitude exhibits a sharp maximum very close to the temperature where the harmonic phase reaches its minimum, T$_m$, then drops abruptly to the noisefloor. In the more underdoped samples, the maximum of the harmonic voltage magnitude occurs at a temperature lower than $T_m$, thus in the superconducting state, but the harmonic response extends above $T_m$ into the pseudogap phase. This doping-dependent nonlinear response could be due to enhanced Cooper pair lifetime \cite{Tan,Mishonov-PRB}, to superconducting fluctuations in the pseudogap phase of underdoped cuprates \cite{Anderson,Dorsey,VarlamovReggiani,Puica} or perhaps to a transition from a pure $d$-wave order parameter to a $d+s$ order parameter \cite{Kurin}. Another source of nonlinear response above T$_c$ could be vortex-like excitations in the pseudogap phase \cite{Golovashkin}.

By a proper choice of the complex conductivity $\sigma_{1,2}$ and its nonlinear corrections $A_{1,2}$, one could incorporate microscopic details (symmetry of the order parameter, shape of the Fermi surface, effects due to the quasiparticles at the nodes of the gap, etc.), anisotropy effects of the in-plane conductivity and various possible sources of nonlinear behavior, such as vortex motion, weak links and defects due to sample fabrication and annealing. Further investigations of nonlocal electrodynamics \cite{Oates2008} should also be explored along with more microscopic models.

\section{
CONCLUSIONS
}\label{conclusions}

A new experimental technique is presented where a vector network analyzer in frequency offset mode is used to acquire the harmonic nonlinear response of homogeneous thin superconducting films to microwave current excitation. The phase-sensitive harmonic detection technique provides an additional piece of information, compared to previous investigations of nonlinear effects in the superconducting state: the phase of the harmonic voltage at temperatures close to T$_c$. The third-order harmonic phase gradually decreases from $\pi/2$ in the superconducting state, reaches a minimum close to T$_c$ and recovers back to $\pi/2$, at least for samples near optimal doping. In the underdoped samples the phase does not recover completely to $\pi/2$ in the normal state, but to roughly 0.5 radians. In all samples used in this study the harmonic magnitude exhibits a maximum, as observed by other investigators \cite{Pestov, Lee-PRB1} and the magnitude maximum occurs at a temperature $T_M$ below that associated with the minimum of phase, $T_m$. A consistent trend with oxygen doping has been found, where the difference $\Delta T_{M,m}=T_m-T_M$ increases in the more oxygen-deficient samples.

An analytical finite-frequency field-based model of the nonlinear microwave response of superconducting thin films in a near-field microwave experimental configuration is presented. The interplay of inductive and resistive nonlinear effects arises naturally in the model, being an improvement with respect to previous models from the literature, which treat the two types of nonlinear behavior separately. The description is field-based as opposed to lumped-element-based and it introduces the nonlinear effects in a phenomenological fashion as deviations of conductivity from its linear-response value. The new model is in agreement with experimental data and other models from the literature in the limiting case of low temperature where the field screening is due to the superfluid, and reproduces some key features observed in the data acquired with the new apparatus. More specifically, at low temperatures it predicts a harmonic phase of $\pi/2$ with a nonlinear behavior originating mainly from the $\sigma_2$ nonlinearity. The model also shows, in agreement with our experimental data, that as temperature increases toward T$_c$ an in-phase, resistive-like, component becomes significant, thus "rotating" the third-order harmonic voltage clockwise in the complex plane. In the regime where the superconducting order parameter relaxation time diverges, the counter-clockwise rotation of the harmonic voltage in the complex plane can be qualitatively reproduced by a phenomenological modification of the model.

We acknowledge useful conversations on nonlinear phase measurements with Mario Mule and O. J. Danzy from Agilent and for the analytical model with Dr. Anatoly Utkin from the Institute of Physics of Microstructures of the Russian Academy of Sciences, Nizhny Novgorod, Russia. The authors wish to acknowledge the support of the National Science Foundation NSF-GOALI DMR-0201261.

\end{document}